\documentclass[showpacs,preprintnumbers,twocolumn,10pt]{revtex4}
\usepackage{amssymb}
\usepackage[dvips]{graphicx}
\usepackage[latin2]{inputenc}
\usepackage[T1]{fontenc}

\begin{document}
\title{Flux qubit on mesoscopic nonsuperconducting ring}
\author{E. Zipper, M. Kurpas, M. Szel\k{a}g, J. Dajka and M. Szopa}
\affiliation{Instytut Fizyki, Uniwersytet \'{S}l\k{a}ski, 40-007  Katowice, Poland}
\begin{abstract}
The possibility of making a flux qubit on nonsuperconducting mesoscopic
ballistic quasi 1D ring is discussed. We showed that such ring can be
effectively reduced to a two-state system with two external control
parameters. The two states carry opposite persistent currents and are
coupled by tunneling which leads to a quantum superposition of states. The
qubit states can be manipulated by resonant microwave pulses. The flux state
of the sample can be measured by a SQUID magnetometer. Two or more qubits
can be coupled by the flux the circulating currents generate. The problem of decoherence is also discussed.
\end{abstract}
\pacs{03.67.Lx, 74.50+r, 73.23.Ra}
\keywords{}
\maketitle
\section{Introduction}
Recently a number of systems which can be effectively reduced to a two-level
systems [1-4] have been examined as candidates for quantum computing
hardware. These include e.g. ion traps, nuclear spins in molecules, charge and flux states of superconducting circuits. Maintaining the coherence of quantum device is a major challenge - the device should be maximally decoupled from the environment to avoid decoherence and thus the loss of quantum information. In addition to further work on existing systems new candidates for qubits can be investigated.

We are considering the mesoscopic ring made of a metal, semiconductor or toroidal carbon nanotube.
Persistent currents (PC) in small nonsuperconducting rings threaded by a magnetic
flux are a manifestation of quantum coherence in a submicron system. If the
ring circumference $L$ is smaller than the phase coherence length $L_{\phi }$
the electron wave function may extend coherently over $L$ even in the
presence of elastic scatterers [5,6]. In other words a normal loop with $%
L<L_{\phi }$ has a nontrivial ground state with a circulating PC.

The goal of this paper is to show that quantum tunneling between states with
nearly equal energies and opposite persistent currents in clean
mesoscopic nonsuperconducting quasi 1D ring with a barrier can lead to a formation of a qubit.
We argue that at low T such ring can be
effectively reduced to a two-state system with two external control
parameters. Quantum tunneling between the states leads to the quantum superposition of two opposed current states. The problem of the qubit decoherence is also discussed.

\section{Mesoscopic normal ring as a two state system - a formation of a
qubit}

Let us consider the mesoscopic metallic or semiconducting quasi $1D$ ring of
radius $R$ ($2\pi R<L_{\phi })$ in the presence of static magnetic flux $%
\phi _{e}$, $\phi _{e}=B_{e}\pi R^{2}$, $B_{e}$ is the applied
magnetic field perpendicular to the plane of the ring. Mesoscopic systems
are expected to behave according to the laws of quantum mechanics if they
are separated well from the external degrees of freedom. Thus we assume that
the system is well insulated from the environment.


We assume that a ring is made from a very clean material i.e. we are in the
ballistic regime. 
The
 energy levels  in a quasi $1D$ ring
are:
\begin{equation}
E_{n}^{0}=\frac{\hbar ^{2}}{2mR^{2}}\left( n-\phi ^{^{\prime }}\right) ^{2},%
\label{jeden}
\end{equation}%
where $\phi ^{^{\prime }}=\frac{\phi }{\phi _{0}}$, $\phi _{0}=\frac{h}{e}$, 
$n=0,\pm 1,...$ is the orbital quantum number (winding number) for an
electron going around the ring. 
With each energy level we can associate a current 
\begin{equation}
I_{n}=-\frac{\partial E_{n}^{0}}{\partial \phi }=\frac{e\hbar }{2\pi mR^{2}}%
\left( n-\phi ^{^{\prime }}\right) ,\text{\quad }n=0,\pm 1,...  \label{dwa}
\end{equation}%
The current is persistent at $kT\ll \Delta $.
In calculating the flux $\phi $ for very
thin ring the selfinductance effect can be neglected and $\phi =\phi _{e}$.
The system has a set of quantum size energy gaps, the gap at the Fermi
surface (FS) at $\phi ^{^{\prime }}=0$ is $\Delta =\frac{hv_{F}}{L}$, where $%
v_{F}$ is the electron velocity at the FS. Assuming the radius $R=400%
\mathring{A}$, we get $\Delta \sim 290K$ for metallic ring and $\Delta
\thicksim 47K$ for semiconducting ring \cite{mailly}.
\begin{figure}
\includegraphics[width=3.2283in, height=2.2641in]{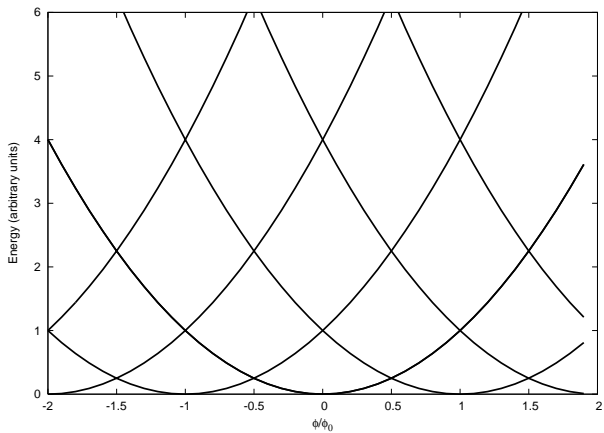}
\caption{The energy spectrum of a 1D ring as a function of the flux $\protect\phi ^{^{\prime }}.$}
\end{figure}

In the following we assume that the temperatures are close to zero and the system behaves coherently (the
energy gap hampers the inelastic transitions).
 The energy spectrum as a
function of $\phi ^{^{\prime }}$ is shown in FIG.~1. Neglecting the spin,
each level $E_{n}$ is occupied by a single electron. We can see from FIG.~1
that at $\phi ^{^{\prime }}=0$ (and all integral $\phi ^{^{\prime }}$), if
the number of electrons in the ring $N=N_{even}=2n_{F}$ then the level at
the FS is doubly degenerate and occupied by a single electron only. The same
situation happens at half integral $\phi ^{^{\prime }}$, for $%
N=N_{odd}=2n_{F}+1$. Because the energy spectrum is periodic with period $%
\phi ^{^{\prime }}=1$, we can restrict our considerations to the
neighborhood of the two degeneracy points $\phi ^{^{\prime }}=0$ (for $%
N_{even})$ and $\phi ^{^{\prime }}=\frac{1}{2}$\ (for $N_{odd})$. The
electron at the FS behaves there as a particle in a double well potential,
where the states in each well correspond to PC of opposite sign. It follows
from (\ref{dwa}) that with increasing magnetic field the ground state will change
from angular momentum $n_{F}$\ to one with higher $n$.

We now introduce to the ring an energy barrier of finite length $a<L$ and height $V$
(positive or negative). In this case the tunneling occurs which mixes the states from both
sides of the barrier. At the degeneracy points the eigenstates which are superposition of states with
different winding numbers can be formed. This causes the splitting of the
initial energy levels (FIG.~2). Quantum tunneling should thus lead to a
qubit i.e. a quantum superposition of the two opposed current states. The
second quantized Hamiltonian in presence of the barrier is%
\begin{equation}
H=\sum\limits_{m\neq n}\left[ E_{n}\left\vert n\right\rangle \left\langle
n\right\vert -\frac{1}{2}\hbar (\omega _{m,n}\left\vert n\right\rangle
\left\langle m\right\vert +\omega _{m,n}\left\vert m\right\rangle
\left\langle n\right\vert )\right]  \label{osiemnascie}
\end{equation}%
where $E_{n}=E_{n}^{0}+V_{n},$ $V_{n}$ is the renormalization due to the
barrier of the $E_{n}^{0}$ at a given $\phi ^{^{\prime }}$, $\omega _{m,n}$
is the phase slip rate between states $\left\vert m\right\rangle $ and $%
\left\vert n\right\rangle $. 

At $\phi ^{^{\prime }}=0$ and at $kT\ll
\Delta $ the energy states with $\left\vert n\right\vert <n_{F}$ are fully
occupied and form the "Fermi sea". The energy states for $\left\vert
n\right\vert >n_{F}$ are separated by large energy gaps from the FS and are
fully empty. The same separation takes place at $\phi ^{^{\prime }}=\frac{1}{%
2}$. Thus the only states which can take part in the tunneling are the
states in the immediate neighborhood of the FS and we can consider a
mesoscopic ring as a two-state quantum system.

In this case the summation in (\ref{osiemnascie}) can be restricted to two
states closest to the FS. If we assume that $N=N_{even}$ and $\phi^{^{\prime }}$is close to $0$ these states are
$\left\vert n_{F}\right\rangle =\left\vert \beta \right\rangle ={1 \choose 0}$ 
and $\left\vert -n_{F}\right\rangle $ $=\left\vert \alpha \right\rangle ={0 \choose 1}$ 
and the Hamiltonian (\ref{osiemnascie}) becomes%
\begin{equation}
H=\left[ 
\begin{array}{cc}
E_{\alpha } & -\frac{1}{2}\hbar \omega _{\alpha ,\beta } \\ 
-\frac{1}{2}\hbar \omega _{\alpha ,\beta } & E_{\beta }%
\end{array}%
\right]   \label{dziewietnascie}
\end{equation}
where $E_{\alpha }=E_{-n_{F}},E_{\beta }=E_{n_{F},}\omega _{\alpha ,\beta
}=\omega _{-n_{F}},_{+n_{F}}.$

For most values of $\phi ^{^{\prime }}$ the $\omega _{\alpha ,\beta }$ is
small compared with the energy of orbital motion of an electron in the ring.
However close to the degeneracy points the $\hbar \omega _{\alpha ,\beta }$
term mixes the two states strongly. At\ $\phi ^{^{\prime }}=$ $0$ the states 
$\left\vert \beta \right\rangle $ and $\left\vert \alpha \right\rangle $
have exactly the same energies $E_{\alpha }=E_{\beta },$ but opposite
currents $I_{\alpha }=-I_{\beta }.$ In this case the phase slip probability
increases (it happens at all integer $\phi ^{^{\prime }}$).

In case of $\phi ^{^{\prime }}=\frac{1}{2}$ we find $\alpha =-n_{F}$ and $%
\beta =n_{F}+1$ i.e. the energies involved are$\ E_{\alpha }=E_{-n_{F}},\
E_{\beta }=E_{n_{F}+1}$ and $\omega _{\alpha ,\beta }=\omega
_{-n_{F}},_{n_{F}+1}$. These states are degenerate and the respective
currents are opposite. Notice, that the case with $N=N_{even}$ and $\phi
^{^{\prime }}=0$ does not require any external field and thus may be easier
to decouple from the environment.

The amplitude of these currents is\ $I_{0}=\frac{ev_{F}}{L}$. Assuming\ e.g. 
$R=400\mathring{A}$ \ $I_{0}\thicksim 1$ $\mu A$ for metallic ring, $%
I_{0}\thicksim 0.16$ $\mu A$\ \ for semiconducting ring.

In a pseudospin notation (\ref{dziewietnascie}) can be written as
\begin{equation}
H=-\frac{1}{2}B_{z}\widehat{\sigma }_{z}-\frac{1}{2}B_{x}\widehat{\sigma }%
_{x}\text{.}  \label{Hsigma}
\end{equation}%
where $\widehat{\sigma }_{z},\widehat{\sigma }_{x}$ denote Pauli spin
matrices. The term $B_{z}$ can be tuned by the applied flux

\begin{equation}
B_{z}=E_{\beta }-E_{\alpha }=\left\{ 
\begin{tabular}{ll}
$\Delta \left[ 1-2\frac{\phi }{\phi _{0}}\right] $ & for $N=N_{odd}$ \\ 
$-\Delta \cdot 2\frac{\phi }{\phi _{0}}$ & for $N=N_{even}$%
\end{tabular}%
\right.  \label{dwanascie}
\end{equation}%
The $x$ component of the effective magnetic field $B_{x}$ describes the
tunneling amplitude $\hbar \omega _{\alpha ,\beta }$ between the two
potential wells and can be tuned by e.g. electrical gating. With these two
external control parameters the elementary single-bit operations i.e. $z$
and $x$ rotations \cite{makh} can be performed. The qubit can be driven by
microwave pulses. The advantage of the proposed qubit is the large distance $%
\Delta $ between the qubit energy levels and the next higher states.

Diagonalizing the hamiltonian (\ref{dziewietnascie}) we obtain two energy
bands

\begin{equation}
E_{\pm }=\pm \frac{1}{2}\sqrt{(E_{\beta }-E_{\alpha })^{2}+\hbar ^{2}\omega
_{\alpha ,\beta }^{2}}.  \label{trzynascie}
\end{equation}%

At the degeneracy point$\ E_{\beta }=E_{\alpha },$
the energy splitting can be estimated by making use of a transfer matrix method \cite{gogolin}. One obtains
\begin{equation}
\hbar \omega _{\alpha ,\beta }=\frac{\Delta }{\pi }\arccos \sqrt{T_{F}}\cos
\phi ^{^{\prime }}\left( 1-\frac{\delta _{F}}{\pi N}\right)  \label{mnascie}
\end{equation}%
where $T_{F},\delta _{F}$ are the transmission probability and the phase
shift of an electron at the FS respectively. The energy eigenstates are then the symmetric and antisymmetric superpositions of states with opposite PC.

We performed numerical calculations of
the energy levels $E_{\pm }$ for different sets of the barrier parameters
and extracted the magnitude of $\hbar \omega _{\alpha ,\beta }$ from it. The
results are shown in FIG.~2 for a ring with $\ N_{even}$, $n_{F}=20$\ \
near $\phi ^{^{\prime }}=0$ for $V=0.005 eV$, $E_{F}\thicksim 0.01eV, a=15%
\mathring{A}$. The initial energy levels are splitted and shifted by the
presence of the barrier.
\begin{figure}
\begin{center}
\includegraphics[width=3.2283in, height=2.2641in]{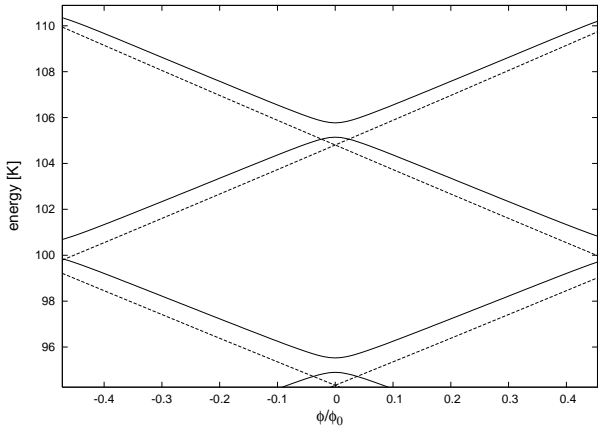}
\caption{The energy levels $E_{\pm }$\ as a function of the flux\ $\protect\phi %
^{^{\prime }}$for R=400\AA\ and$\ N_{even}$ with $n_{F}=20$ (solid lines). Superposition
states manifest themselves as an anticrossing of the initial energy levels (dashed lines).}
\end{center}
\end{figure}
The qubit energy splitting depends on the width and height of the
barrier which can be raised or lowered by electrical gating. It is shown in
FIG.~3.
\begin{figure}
\begin{center}
\includegraphics[width=3.2283in, height=2.2641in]{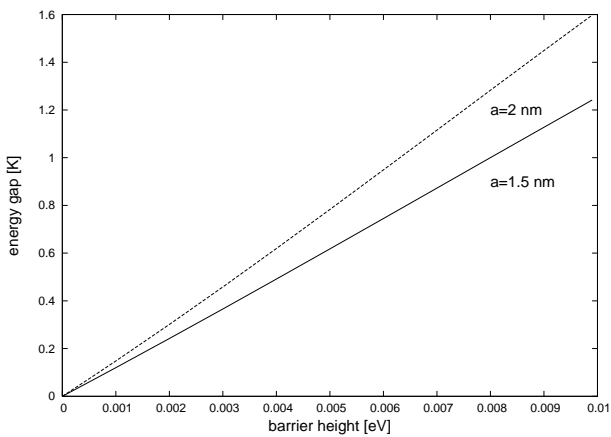}
\caption{Qubit energy splitting $\hbar \omega _{\alpha ,\beta }$ at $\phi ^{^{\prime }}=0$ as a function of the barrier
height $V$ for two values of the barrier width a.}
\end{center}
\end{figure}

The potential barrier in a mesoscopic ring can be realized in a number of
ways. One of them is the ring in which the small fragment of a convex arc
have been deformed to form a concave one \cite{pershin}. One can produce
such a mechanical deformation by pressing. Another possibility to form the
barrier is the local point-like electric gate close to the ring. The
electric potential of the gate forms the barrier, the height of which can be
modified by the gate voltage. The electric potential barrier can be combined
with the geometric one described above. In this case the barrier height and
therefore the mixing term $\omega _{\alpha ,\beta }$ can be adjusted to the
required value. To make the barrier more local we can use Atomic Force Microscope or Scanning Tunneling Microscope. Additionally
one can regulate the barrier height by the change of the potential of the
tip. The other possibility is to place the single wall carbon nanotube in
the plane of the ring perpendicular to it and apply the voltage to it.

Another realization of such system can be obtained by using a toroidal
carbon nanotube (CN). It is known that the defect free metallic CN behaves
like real 1D wire \cite{latil}. This is due to the small (two) number of
channels available close to the FS. The FS consists of two points where the
two bands cross. The properties of an ideal toroidal CN depend on the
position with respect to the honeycomb lattice of two circles: circumference
of CN and circumference of the torus, defining chiral and twist vectors $%
(p_{1},p_{2};q_{1},q_{2})$ respectively \cite{marszo}. In our model we
assume that only one electron occupies the FS which is doubly degenerate. In
CN-s which are metallic in both vectors: $p_{1}-p_{2}=3k$ and $%
q_{1}-q_{2}=3l $, where $k$ and $l$ are integers, there are four accessible
states at the FS. If however a CN is metallic only in the chiral direction $%
p_{1}-p_{2}=3k$ and not metallic in twist direction $q_{1}-q_{2}\neq 3l$, we
obtain the required structure by applying a flux $\phi ^{^{\prime }}=\pm 
\frac{1}{3}+integer$, in the direction parallel to the torus symmetry
axis. The dispersion relation of CN \cite{dres, marg} is different from that
of quasi 1D ring. However, in the neighborhood of the Fermi points both
relations become similar i.e. linear in $\phi ^{^{\prime }}$ and producing
opposite currents \cite{minot}. Because the states close to the FS are well
separated from the excited states the system can also be treated as a
two-level system. As an example we consider the metallic armchair CN ($%
10,10;-5000,5000 $) having the radius $R~\thicksim~1960\mathring{A}$ and the width $d=8\mathring{A}$ \cite{lin}. The energy gap between the
first unoccupied state and the state at the FS is $\Delta \thicksim 32K$.
The current of an electron at the FS can be roughly approximated by

\begin{equation}
I_{n_{F}}\thicksim I_{0}\left( 1-\phi ^{^{\prime }}\right) ,\text{ \ \ }%
0\leq \phi ^{^{\prime }}\leq 1 
\end{equation}
The current amplitude$\ I_{0}$ is inversely proportional to $R$ but
independent of the toroid width. One obtains $I_{0}\thicksim 0.54 \mu A$ or equivalently
the magnetic flux $\phi \thicksim 0.2\cdot 10^{-3}\phi _{0}$ for the assumed
parameters.

The mixing term $\omega _{\alpha ,\beta }$ is again due to the potential
barrier along the toroidal CN. It can be obtained by a non smooth
junction of both its ends by e.g. a fulleren molecule \cite{joa} or by simply leaving a small gap between the ends.
Alternatively, one can replace a fragment of a toroidal nanotube with the
one of similar circumference but different conductivity properties (e.g. $%
\left( 11,0\right) $ and $\left( 6,6\right) $) \cite{jaskolski}. Finally, to
obtain a system described by (\ref{Hsigma}) one can apply a real magnetic
field perpendicular to the torus symmetry axis close to the degeneracy point 
$(B_{z}=0)$ \cite{tsai}$.$

In our model calculations we have made simplifying assumptions that the ring
is quasi $1D$ and its energy states (\ref{dwa}) are single particle states.
However the presented considerations are also valid with some modifications\
for the mesoscopic metallic or semiconducting ring with very small thickness 
$d$, $d\ll R$ \ i.e. with a few transverse channels \cite{mailly}, \cite{fuhrer}.
The nanoscopic, semiconducting, defect-free quasi $1D$ rings in which the
electronic states are in the true quantum limit have been already realized 
\cite{lorke}. The change in the ground state angular momentum numbered by $n$
has been observed in a magnetic field perpendicular to the plane of the
ring. They also found that the single particle states are a quite accurate
basis for a description of the many particle states. Similar findings have
been also obtained for CN tori \cite{latil}, \cite{minot}. It is also
supported by persistent current measurements \cite{mailly} on ballistic
semiconducting rings with a few transverse channels. It was found that the
measured current is of the same order as the current amplitude of
noninteracting electrons $I_{0}$. This means that electron-electron
interaction in real quantum rings is not so strong. For weakly interacting
electrons we can consider the effect of the interaction \cite{krive} by a
barrier renormalization. The electron is scattered not only by the barrier but also by the potential
induced by charge density fluctuations.

The renormalized transmission probability $T_{F}^{R}$ at \thinspace $T\ll~\Delta $ is
\begin{equation}
T_{F}^{R}=\frac{T_{F}\left( \frac{\Delta }{E _{F}}\right) ^{2\xi }%
}{\left( 1-T_{F}\right) +T_{F}\left( \frac{\Delta }{E _{F}}\right)
^{2\xi }}
\end{equation}

where $\xi =\frac{w}{hv_{F}}$ characterizes the strength of the electron -
electron interaction, $w$ is the forward scattering amplitude of the
interaction. The interaction parameter $\xi$ can be expressed in terms
of the Luttinger liquid stiffness constant $\alpha =\frac{v_{F}}{s}$ ($s$
is the plasmon velocity):
\begin{equation}
\xi =\frac{1}{2}\left( \frac{1}{\alpha ^{2}}-1\right) .
\end{equation}
For weakly interacting electrons ($\alpha \rightarrow 1$) $\xi \sim \alpha ^{-1}-1$ and the effect of
the interactions is small. 

Thus the idea of the formation of flux qubit on mesoscopic ring is still
valid for the interacting electrons.

Till now we neglected the electron spin. The orbital magnetic moments \cite%
{minot}, \cite{fuhrer} in small ring structures are an order of magnitude
larger than spin moments and usually the orbital states are successively
populated with spin-up and spin-down electrons. Neglecting the small spin
splitting in the magnetic field our picture with spin included does not
change qualitatively but more possibilities occur. If the number of electron
is $N=2\left( 2n_{F}+1\right) $ the tunneling can take place at $\phi
^{^{\prime }}=\frac{1}{2}$. For other values of $N$ tunneling can take place
at $\phi ^{^{\prime }}=0$ and/or $\phi ^{^{\prime }}=\frac{1}{2}.$

Flux qubit proposed by us is based on the similar idea as the flux qubit
built on superconducting ring \cite{mooij}. Also the measurement of the flux state can be
performed by a separate SQUID magnetometer inductively coupled to it in a
similar way as in a superconducting case. Two or more qubits can be coupled by means of the flux that the circulating current generate. 

\section{Decoherence}

The quality of  an effective information retrieval   devices depends on the conditions 
of its coherence. The question is to what extent the device behaves quantum mechanically when placed in noisy environment generated    by various fluctuations or measurement.
 Thus the important
constraints on the device are dephasing effects due to various decoherence
sources. Below we discuss and estimate some of the main decoherence sources.

There is a natural question how the typical values of phase coherence length $L_{\phi}$, which is of the order of $10^5 \mathring{A}$ \cite{eckern,mohanty} translate into an applicability of mesoscopic rings as the qubit with relatively long decoherence times.  

We have assumed that $kT\ll~\hbar~\omega _{\alpha ,\beta }\ll~
\Delta $ and that  $L<L_{\phi }$. Under these conditions the currents running
in a state of thermodynamic equilibrium are genuinely persistent \cite{cheung,mohanty,imrybook}. The finite decoherence time of the current is due to the interaction with the outside world and leads to the persistence of the currents on a time scale much longer than the coherence time $\tau_{\phi}\sim L_{\phi}$.

We also assume that a system is put in a shield, that screens it from the unwanted radiation.
Thin mesoscopic quantum  rings are the systems with relatively small number of degrees of freedom comparing with 
other solid state devices based on superconducting rings. However they are still able to accommodate various intrinsic fluctuations \cite{eckern,mohanty,kogan,hangi,imrybook}. 

Thermal motion of any charge carriers is a source of thermal fluctuations  related
to the electronic Nyquist noise. 
At low T the weak electron-phonon coupling gives some decrease of the
current amplitude but it does not lead to substantial level shifting and
broadening. The effect of thermal noise on the equilibrium statistics of persistent currents has been studied in a semi-classical regime in \cite{dajka}.  

The qubit can also decohere by spontaneous emission of photons. It follows from
general considerations that this effect is small for the qubit size smaller
than the radiated wavelength (qubit is then an inefficient antenna)\cite{mooij}. Taking for our qubit e.g. $R=400\mathring{A}$ we estimated $t_{m} \sim 10^9 s$ and for $R=2000\mathring{A}$  $t_{m} \sim 10^8 s$ - thus the radiation is not a
serious source of decoherence. The coupling between the
magnetic moments of the current loops and those of nuclear spins can also be
a reason of decoherence. However it may be considerably reduced by aligning
the spins or by applying the compensating pulse sequences \cite{makh}. We also estimated the dephasing 
from the unwanted dipol-dipol coupling \cite{tian}. It is of the order of $14 ms$ if the qubits are at the distance of $10 \mu m$.
 Because our basic states are flux states, the qubit will be sensitive to a
flux noise but relatively insensitive to a charge noise. Thus we expect that
magnetic degrees of freedom in quantum rings should have longer decoherence
times than charge degrees of freedom. The fluctuations in the barrier and in
the magnetic field necessary to get the degeneracy of the states for $%
N=N_{odd}$ can be also the source of decoherence. However the system for $ N=N_{even}$ is naturally bistable requiring no external bias. 

The most important class of fluctuations comes from the inductive coupling of the qubit to the measuring apparatus which is often a DC SQUID.
It can be analyzed along the treatment developed for superconducting flux qubit \cite{caspar} which can be generalized to related systems. We assume that the measurement is performed with the same device as in \cite{caspar}. The flux qubit coupled to the SQUID is an effective dissipative two state or spin-boson system. There are two time scales related to the effect of environment. The first is the characteristic relaxation time of populations to approach an equilibrium Gibbs-like form. The second is the decoherence time after which coherences become negligible \cite{weiss}.
The relaxation ${\tau_r^{-1}}$ and dephasing ${\tau_\phi^{-1}}$ rates obtained using the spin-boson model  for the system at temperature $T$ are \cite{weiss}
\begin{equation}
 \tau_r^{-1}=\frac{1}{2}(\frac{B_x}{\nu})^2 J(\nu/\hbar)\coth(\frac{\nu}{2k_BT})
 \end{equation}
 and
 \begin{equation}
 \tau_\phi^{-1}=\frac{\tau_r^{-1}}{2}+(\frac{B_z}{\nu})^2\alpha 2\pi \frac{k_B T}{\hbar}
 \end{equation}
 where $J(\omega)$ is a spectral density function characterizing fully the environment and $\alpha=\lim_{\omega\rightarrow 0}J(\omega)/(2\pi\omega)$ is the 'ohmicity' parameter. 
The impedance $Z(\omega)$ of the DC SQUID is a source of thermal voltage fluctuations
\begin{eqnarray}
\langle \delta V\delta V\rangle =\nu \Re \{Z(\omega)\}\coth(\frac{\hbar}{2k_BT})
\end{eqnarray}
of the statistics governed by the Nyquist theorem. This fluctuations are related via current-voltage characteristics to the current fluctuations $\langle \delta I_{sq}\delta I_{sq}\rangle_\omega$ and then inherited by the inductively coupled qubit \cite{caspar}:
\begin{equation}
\delta B_z=-2I\delta\Phi_{sq}
\end{equation}
where $\delta\Phi_{sq}$ is the fluctuation of the magnetic flux due to the measuring device.

 It leads in our case to the following spectral density  
 \begin{equation}
 J(\omega)=\hbar^{-1}(2\pi)^2\frac{1}{\omega}(\frac{MI}{\Phi_0})^2 I^2_{sq}\Re \{Z(\omega)\}
 \end{equation} 
where ${M}$ is the mutual inductance coefficient, $\nu={E_{+}-E_{-}}$ is the qubit level spacing.  
 ${I}$,${I_{sq}}$ are the currents in the qubit and SQUID respectively. The effect of fluctuations on the tunneling term $\sigma_x$ can be included in a similar way. However the effect of such fluctuations is relatively weak. 

In these formulas a dimensionless factor 
\begin{equation}
\gamma=(\frac{MI}{\Phi_0}) 
\end{equation}
is a measure of the coupling of the qubit and the measuring SQUID. Its magnitude 
${\gamma\sim 2\cdot10^{-3}}$ has been estimated  for the superconducting qubit \cite{caspar}. In our case as the size of the nonsuperconducting ring  is smaller, the coupling constant $\gamma$ is also smaller.
For the ring with ${R\sim0.2 \mu m}$ (toroidal nanotube) we estimated ${\gamma\sim 0.9\cdot10^{-3}}$.
Assuming the same parameters for the measuring device as in \cite{caspar} we obtain ${\tau_r\sim75\mu s}$ and ${\tau_{\phi}\sim10^2\mu s}$.

In the nonsuperconducting flux qubit the number of degrees of freedom is relatively small 
in comparison to the superconducting devices. On one hand it limits significantly the number of 
decoherence mechanisms but on the other hand as the system is normal one can not expect the suppression of some fluctuations as in the superconducting case. Therefore because the  state of the mesoscopic ring is not completely stable, some random fluctuations of the current are expected; actually some modifications of the Aharonov-Bohm oscillations on a time scale of $10-40~h$ have been  observed \cite{mailly}.  
Finally, small size of the qubit helps to reduce the influence of the environment \cite{mooij} what is significant for the effective engineering. We can estimate the influence of various decoherence sources but it is impossible to determine the real decoherence time with certainty, except by measurement.

The qubit level spacing sets \cite{makh} the fastest operation time to $\tau _{op}\sim 10~ns$. Thus the quality factor is of the order of $10^4$. Our model could in principle be tested on quantum rings investigated in \cite%
{fuhrer},\cite{lorke} after the application of the effective field $B_{x}$ i.e. when one introduces a controllable potential barrier.

The extended discussion of the effective engineering of the decoherence for the nonsuperconducting flux qubit is postponed to further publication . 

\section{Summary}
The advantage of microscopic quantum systems (atoms, spins) for the qubit
formation is that they can be easily isolated from the environment.
The disadvantage is that the integration of many qubits into complex circuit
is a difficult task. From that point of view solid state devices like charge
and flux qubits built on superconducting rings are easier to integrate in a
quantum computer using standard circuit technology. However the
large number of degrees of freedom makes it more difficult to maintain the
coherence.
The proposed flux qubit built on normal quantum ring is on the border line
between these two structures. The small number of degrees of freedom
together with the small size of the qubit helps to decouple it from the
environment. The proposed qubit can be addressed, manipulated, coupled to
each other and read out.The quality factor giving the number of quantum logic operations is of the order of $10^4$. The proposed qubit should be of considerable interest for fundamental studies of quantum coherence in mesoscopic systems and some aspects of quantum theory such as superposition of quantum states and entanglement.

We thank J. E. Mooij, T. Dietl and M. Marga\'{n}ska for helpful discussions.
Work was supported by Polish Ministry of Scientific Research and Information
Technology under the (solicited) grant No PBZ-MIN-008/P03/2003.
J. D. acknowledges the support of The Foundation for Polish Science under grant SN-302-934/2005

\end{document}